\documentclass[aps,prb,reprint,showpacs]{revtex4-2}
\usepackage[dvipdfmx]{graphicx}
\usepackage{subfigmat}
\usepackage{amsmath,amssymb}
\usepackage{color}
\usepackage{bm}
\usepackage{hyperref}
\usepackage{comment}
\usepackage[normalem]{ulem}

\usepackage{cancel}

\usepackage{upgreek}

\newcommand{\tempinv}{\upbeta}

\begin{document}

\title{Theory of inverse Rashba-Edelstein effect induced by spin pumping \\
into a two-dimensional electron gas}

\author{M. Yama$^{1}$, M. Matsuo$^{2,3,4,5}$, T. Kato$^{1}$,}
\affiliation{
${^1}$Institute for Solid State Physics, The University of Tokyo, Kashiwa 277-8581, Japan\\
${^2}$Kavli Institute for Theoretical Sciences, University of Chinese Academy of Sciences, Beijing 100190, China\\
${^3}$CAS Center for Excellence in Topological Quantum Computation, University of Chinese Academy of Sciences, Beijing 100190, China\\
${^4}$Advanced Science Research Center, Japan Atomic Energy Agency, Tokai 319-1195, Japan\\
${^5}$RIKEN Center for Emergent Matter Science (CEMS), Wako, Saitama 351-0198, Japan\\
}

\date{\today}

\begin{abstract}
The inverse Rashba-Edelstein effect (IREE) in a two-dimensional electron gas (2DEG) induced by spin pumping from an adjacent ferromagnetic insulator (FI) is investigated theoretically. In particular, spin and current densities in the 2DEG in which both Rashba and Dresselhaus spin-orbit interactions coexist are formulated, and their dependencies on ferromagnetic resonance frequency and orientation of the spin in the FI are clarified. It is shown that spin density diverges when the ratio between the Rashba and Dresselhaus spin-orbit interactions approaches unity, while current density stays finite there. These results can be applied for evaluating spin splitting on the Fermi surface in a 2DEG and designing spintronic devices using IREE.
\end{abstract}
\maketitle 

\section{Introduction}
\label{sec:introduction}

The conversion phenomenon from charge current to spin polarization in a system without spatial inversion symmetry is called the Rashba-Edelstein effect (REE)\cite{Aronov1989, Aronov1991, Edelstein1990, Inoue2003, Silsbee2004, Sinova2015, Soumyanarayanan2016}. REE, which is also known as the inverse spin-galvanic effect\cite{Gambardella2011, Manchon2019}, in a two-dimensional electron gas (2DEG) with Rashba spin-orbit interaction has been extensively studied \cite{Bychkov1984, Rashba2015, Winkler2007, Manchon2015}. The inverse conversion from spin polarization to charge current is called the inverse Rashba-Edelstein effect (IREE)\cite{Sanchez2013, Shen2014a, Soumyanarayanan2016} (or the spin-galvanic effect\cite{Ganichev2002, Ivchenko1989, Ivchenko1990, Ganichev2001, Ganichev2003b, Burkov2004, Sinova2015}). Both REE and IREE are now important phenomena in the field of semiconductor spintronics\cite{Fabian2007, Awschalom2007, Manchon2015, Kohda2017, Dieny2020}.

Recently, regarding development of spintronic devices, spin-charge conversion combining REE or IREE with conventional methods of spintronics has been attracting much attention. For example, ferromagnetic resonance (FMR) has been used to inject electron spins into a target system from an adjacent ferromagnet. Combined with IREE, this technique, called “spin pumping,” has been used ~\cite{Tserkovnyak2002, Tserkovnyak2005, Hellman2017} to generate charge current in materials such as Ag/Bi\cite{Sanchez2013, Nomura2015, Sangiao2015, Zhang2015, Matsushima2017}, STO\cite{Lesne2016, Song2017, Vaz2019, Noel2020, Ohya2020, Bruneel2020, To2021, Trier2022}, topological insulators\cite{Shiomi2014, Sanchez2016, Wang2016, Song2016, Mendes2017, Sun2019, Singh2020, Dey2021, He2021, Zhang2016}, atomic layers\cite{Mendes2015, Dushenko2016, Mendes2019, Bangar2022}, and semiconductors\cite{Chen2016, Oyarzun2016}. 
Semiconductors with the zinc-blende structure exhibit two kinds of spin-orbit interactions, namely, Rashba ones and Dresselhaus ones\cite{Dresselhaus1955, Rocca1988, Winkler2007}. These spin-orbit interactions cause spin-dependent transport phenomena, such as the Aharonov-Casher effect\cite{Nagasawa2018}, and exhibit the persistent spin helix (PSH) state\cite{Bernevig2006, Weber2007, Koralek2009, Kohda2012, Sasaki2014} when they compete with each other. REE and IREE in a 2DEG in the presence of these two types of spin-orbit interactions have been experimentally investigated widely\cite{Ganichev2003a, Ganichev2003b, Ganichev2004, Giglberger2007, Belkov2008, Ganichev2008, Ganichev2014, Sheikhabadi2017, Tao2021, Zhuravlev2022} and theoretically analyzed by using the Boltzmann or Eilenberger equations\cite{Shytov2006, Raimondi2006, Trushin2007, Raichev2007, Engel2007, Gorini2010, Raimondi2012, Xintao2013, Shen2014b, Gorini2017, Sheikhabadi2018, Tao2021, Tkach2021, Tkach2022, Suzuki2023}. Recently, IREE combined with spin pumping has begun to be studied theoretically\cite{Tolle2017,Dey2018,Fleury2023}. 
In these works, spin-orbit interactions are assumed to be much weaker than energy broadening due to impurity scattering.  
However, as for a clean 2DEG formed at a semiconductor interface, the opposite case, namely, impurity strength is weaker, is frequently encountered\cite{Umansky1997}. 

\begin{figure}[tb]
  \begin{center}
  \includegraphics[width=80mm]{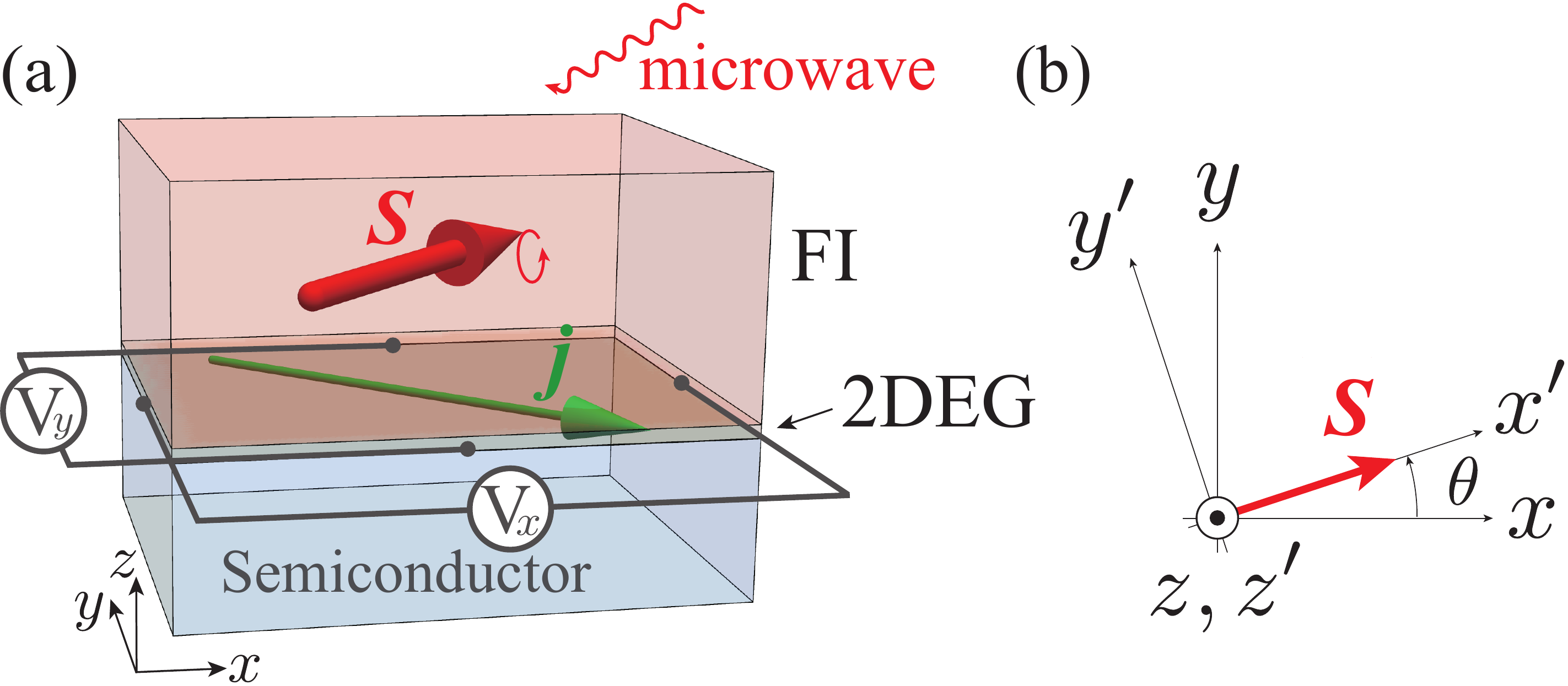}
     \caption{(a) Setup for studying the inverse Rashba-Edelstein effect (IREE) induced by spin pumping. 
     The red arrow, $\bm{S}$, indicates spontaneous spin polarization of the ferromagnetic insulator (FI), while the green arrow, $\bm{j}$, represents current density generated by IREE. (b) Relation between laboratory coordinates $(x,y,z)$ and magnetization-fixed coordinates $(x',y',z')$. 
     The red arrow indicates $\bm{S}$, i.e., the spontaneous  spin polarization of the FI.}
    \label{fig:setup}
  \end{center}
\end{figure}

In this study, as shown in Fig.~\ref{fig:setup}~(a), IREE in a 2DEG induced by spin pumping from an adjacent ferromagnetic insulator (FI) is considered. 
The Boltzmann equation is used to clarify the dependences of spin and current densities produced by IREE on FMR frequency and orientation of the spin polarization in the FI [see Fig.~\ref{fig:setup}~(b)]. 
It is shown that the current induced by IREE includes information of spin texture near the Fermi surface. The influence of the ratio between the Rashba and Dresselhaus spin-orbit interactions on the maxima of spin and current densities is also clarified. These results can be applied to interfacial 2DEG systems coupled with an FI, which can be formed in, e.g., YIG/GaAs/AlGaAs and YIG/GaAs junctions. The experimental feasibility will be discussed in more detail in Sec.~\ref{Sec:Experiment}.

In this study, we focus on the weak-impurity case, namely, the spin-orbit interactions in the 2DEG are much larger than energy broadening due to impurity scattering, while they are much smaller than the Fermi energy. It should be noted that in this situation either the Dyakonov-Perel (DP) or Elliot-Yafet (EY) mechanisms do not hold\cite{Szolnoki2017, Suzuki2023} and that spin currents in the 2DEG are no longer well-defined\cite{Khaetskii2006, Shitade2022}. Accordingly, we formulate the IREE in the 2DEG without using a spin current\cite{Suzuki2023}.

The rest of this work is organized as follows. Model Hamiltonians of the 2DEG/FI bilayer system are presented in Sec.~\ref{sec:model}. Spin and current densities in the 2DEG induced by IREE are calculated using the Boltzmann equation in Sec.~\ref{sec:formulation}. In Sec.~\ref{sec:results}, the spin and current densities are plotted as functions of FMR frequency and orientation of the spin polarization in the FI. In Sec.~\ref{Sec:Experiment}, the experimental feasibility of our results is discussed. The results of this study are summarized in Sec.~\ref{sec:summary}.

\section{Model}
\label{sec:model}

A microscopic model for describing the 2DEG/FI junction shown in Fig.~\ref{fig:setup} is introduced. The Hamiltonians for a 2DEG (Sec.~\ref{sec:2DEG}), a FI (Sec.~\ref{sec:FI}), and interfacial coupling between the 2DEG and FI (Sec.~ \ref{sec:interface}), are described in that order.

\subsection{Two-dimensional electron gas}
\label{sec:2DEG}

A second-quantized Hamiltonian of a 2DEG with both Rashba and Dresselhaus spin-orbit interactions is written as
\begin{align}
& H_{\rm kin}=\sum_{\bm{k}}
\begin{pmatrix}
c^{\dagger}_{\bm{k}\uparrow} & c^{\dagger}_{\bm{k}\downarrow} 
\end{pmatrix}
\hat{h}_{\bm{k}}
\begin{pmatrix}
c_{\bm{k}\uparrow} \\ c_{\bm{k}\downarrow} 
\end{pmatrix},\\
& \hat{h}_{\bm{k}}=
(\epsilon_{\bm k} -\mu)\hat{I}+\alpha(k_y\hat{\sigma}_x-k_x\hat{\sigma}_y)+\beta(k_x\hat{\sigma}_x-k_y\hat{\sigma}_y),
\end{align}
where $c_{{\bm k}\sigma}^\dagger$ ($c_{{\bm k}\sigma}$) is the creation (annihilation) operator of an electron with wavenumber $\bm{k}=(k_x,k_y)$ and spin $\sigma$ ($=\uparrow, \downarrow$), $\epsilon_{\bm{k}} = \hbar^2(k_x^2+k_y^2)/2m^*$ is energy dispersion of conduction electrons, $m^*$ is effective mass of conduction electrons, and $\mu$ is chemical potential. The Fermi energy $\epsilon_{\rm F}$ is defined as the chemical potential at zero temperature, and the Fermi wavenumber is defined as $\epsilon_{\rm F}=\hbar^2 k_{\rm F}^2/2m^*$. The $2\times 2$ matrix $\hat{h}_{\bm k}$ is written with identity matrix $\hat{I}$ and Pauli matrices $\hat{\bm \sigma}=(\hat{\sigma}_x, \hat{\sigma}_y)^{T}$. The amplitudes of the Rashba and Dresselhaus spin-orbit interactions are denoted with $\alpha$ and $\beta$, respectively. $\hat{h}_{\bm{k}}$ can be rewritten in terms of effective Zeeman field $\bm{h}_{\rm eff}=(h_x,h_y)^T$ as
\begin{align}
\hat{h}_{\bm{k}}&= (\epsilon_{\bm k}-\mu)\hat{I}-\bm{h}_{\rm eff}\cdot\hat{\bm{\sigma}},\\
\bm{h}_{\rm eff}(\bm{k})
&=|\bm{k}|\left(\begin{array}{c}
-\alpha\sin\varphi-\beta\cos\varphi \\
\alpha\cos\varphi+\beta\sin\varphi \end{array} \right),
\label{eq:Zeeman}
\end{align}
where $\bm{k}=(|\bm{k}|\cos\varphi,|\bm{k}|\sin\varphi)$. When spin-splitting in band dispersion is calculated, it is assumed that the spin-orbit interaction energies, i.e., $k_{\rm F}\alpha$ and $k_{\rm F}\beta$, are much smaller than the Fermi energy and approximated as
\begin{align}
\bm{h}_{\rm eff}(\bm{k})
&\simeq k_{\rm F}\left(\begin{array}{c}-\alpha\sin\varphi-\beta\cos\varphi \\ \alpha\cos\varphi+\beta\sin\varphi \end{array} \right) , 
\label{eq:Zeeman2}
\end{align}
It follows that ${\bm h}_{\rm eff}$ depends only on azimuth angle $\varphi$ and can be denoted by ${\bm h}_{\rm eff}(\varphi)$. The conduction band is split into two spin-polarized bands, whose energy dispersion is given as
\begin{align}
E^{\gamma}_{\bm{k}}&=\epsilon_{\bm{k}}+\gamma h_{\rm eff}(\varphi), \label{eq:Egk}\\
h_{\rm eff}(\varphi)&\equiv |\bm{h}_{\rm eff}(\varphi)|=k_{\rm F}\sqrt{\alpha^2+\beta^2+2\alpha\beta\sin2\varphi}. \label{eq:def:heff}
\end{align}
where $\gamma$ ($=\pm$) is an index of the spin eigenstate. The corresponding eigenstates are given as
\begin{align}
|\bm{k},\gamma\rangle 
&= \frac{1}{\sqrt{2}}
\begin{pmatrix}
C(\varphi) \\ \gamma
\end{pmatrix},\label{eq:eigenkg}\\
C(\varphi)&\equiv \frac{-h_x(\varphi)+ih_y(\varphi)}{h_{\rm eff}(\varphi)} . \label{eq:defC}
\end{align}
Note that $|\bm{k},\gamma\rangle$ depends only on $\varphi$ and is independent of $|\bm{k}|$. These wavefunctions can be used to introduce the annihilation operator of an electron in the eigenbases as
\begin{align}
c_{\bm{k}\sigma}& =\sum_{\gamma}C_{\sigma\gamma}(\varphi)c_{\bm{k}\gamma}, \\
C_{\uparrow \gamma} &= C(\varphi)/\sqrt{2}, \quad C_{\downarrow \gamma} = \gamma/\sqrt{2}, \label{eq:Cupdown}
\end{align}

\begin{figure*}[tb]
\centering
\includegraphics[width=155mm]{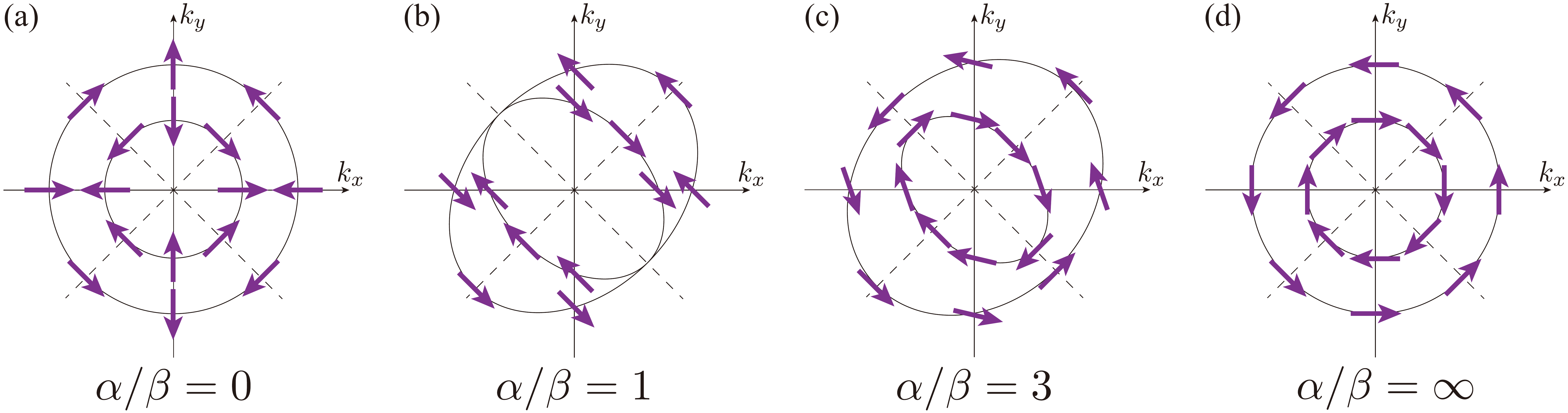}
\caption{Spin texture on the spin-split Fermi surface: (a) $\alpha/\beta=0$, (b) $\alpha/\beta=1$, (c) $\alpha/\beta=3$, and (d) $\alpha/\beta=\infty$. Purple arrows represent spin polarization of 2DEG electrons in each Fermi surface. Note that the spin splitting in the figures is stressed, while the energy splitting is assumed to be much smaller than the Fermi energy in our calculation.
}
\label{fig:spintexture}
\end{figure*}

Spin-split Fermi surfaces for various values of $\alpha/\beta$ are illustrated schematically in Fig.~\ref{fig:spintexture}. As shown in Figs.~\ref{fig:spintexture}~(a) and (d), spin-splitting energy $2h_{\rm eff}$ is constant on the Fermi surface when only the Dresselhaus spin-orbit interaction exists ($\alpha = 0$) or only the Rashba spin-orbit interaction exists ($\beta = 0$). In other cases, $2h_{\rm eff}$ depends on $\varphi$, i.e., the position on the Fermi surface as shown in Figs.~\ref{fig:spintexture}~(b) and (c). The arrows on the Fermi surface indicate the spin polarization of the energy eigenstates.

Electron scattering by nonmagnetic impurities is also considered as follows. The Hamiltonian of the impurities is given as
\begin{align}
H_{\rm imp}&=\sum_l H_{\rm imp}({\bm R}_l), \\
H_{\rm imp}({\bm R}) &=\sum_{\sigma}
\int d\bm{r} \, v(\bm{r}-\bm{R})\psi^{\dagger}_{\sigma}(\bm{r})\psi_{\sigma}(\bm{r}), \\
\psi_{\sigma}(\bm{r})&=\frac{1}{\sqrt{\mathcal{A}}}\sum_{\bm{k}}e^{i\bm{k}\cdot\bm{r}}c_{\bm{k}\sigma},
\end{align}
where $v(\bm{r})$ is the impurity potential, $\bm{R}_l$ is the impurity position at impurity site $l$, and $\mathcal{A}$ is the junction area. For simplicity, a point-like impurity potential is considered as $v(\bm{r})=u\delta({\bm r})$, where $u$ is the strength of the impurity potential. As discussed in Sec.~\ref{sec:formulation}, the effect of impurity scattering is taken into account in the Boltzmann equation in terms of scattering rates, which are calculated by using the Born approximation. The effect of the impurity scattering is represented by energy broadening $\Gamma = 2 \pi n_{\rm imp} u^2 D(\epsilon_{\rm F})$, where $n_{\rm imp}$ is the number density of impurities and $D(\epsilon_{\rm F})$ is the density of states near the Fermi energy. Hereafter, we assume the weak-scattering condition, namely, $\Gamma \ll {\rm max}(\alpha k_{\rm F}, \beta k_{\rm F})$ (for a detail, see Sec.~\ref{sec:formulation}).

\subsection{Ferromagnetic Insulator}
\label{sec:FI}

The quantum Heisenberg model for the FI is considered next. The model is written in laboratory coordinates as
\begin{align}
H_{\rm FI} &= \sum_{\langle i,j \rangle}
J_{ij} {\bm S}_i \cdot {\bm S}_j -\hbar \gamma_{\rm g} \sum_{i} {\bm h}_{\rm dc}\cdot {\bm S}_{i}, \label{eq:HFI1} \\
{\bm h}_{\rm dc} &= (-h_{\rm dc} \cos \theta, -h_{\rm dc} \sin \theta, 0), 
\end{align}
where $J_{ij}$ ($<0$) is the ferromagnetic exchange interaction, $\langle i,j \rangle$ indicates a pair of nearest-neighbor sites, $\gamma_{\rm g}$ ($<0$) is the gyromagnetic ratio, ${\bm h}_{\rm dc}$ is the external static magnetic field, and $\theta$ is the azimuth angle of the magnetic field. The spin-wave approximation is used under the assumption that the temperature is much lower than the magnetic transition temperature and the amplitude of the spin $S_0$ is much larger than unity ($S_0 \gg 1$). The expectation value of the localized spin is then given as $\langle \bm{S}_i\rangle = (\langle S^x_i\rangle, \langle S^y_i\rangle, \langle S^z_i\rangle)=(S_0\cos\theta, S_0\sin\theta, 0)$. For applying the spin-wave approximation, it is convenient to introduce new coordinates $(x',y',z')$, which are fixed in the direction of the ordered spin. In the new coordinates, the expectation value of the spin is expressed as $\langle \bm{S}_i\rangle = (\langle S^{x'}_i\rangle, \langle S^{y'}_i\rangle, \langle S^{z'}_i\rangle)=(S_0,0,0)$ (see Fig.~\ref{fig:setup}~(b)). The spin operators in the two types of coordinates are related to each other as
\begin{align}
\begin{pmatrix}
S^{x'}_i \\ S^{y'}_i \\ S^{z'}_i
\end{pmatrix}
=
\begin{pmatrix}
\cos\theta & \sin\theta & 0
\\ -\sin\theta & \cos\theta & 0 
\\ 0 & 0 & 1
\end{pmatrix}
\begin{pmatrix}
S^{x}_i \\ S^{y}_i \\ S^{z}_i
\end{pmatrix}.
\end{align}
The Holstein-Primakov transformation,
\begin{align}
S^{x'+}_j &= S^{y'}_j + iS^{z'}_j = (2S_0)^{1/2} b_j, \\
S^{x'-}_j &= S^{y'}_j - iS^{z'}_j = (2S_0)^{1/2} b_j^\dagger,\\
S^{x'}_j &= S_0 - b_j^\dagger b_j,
\end{align}
and the Fourier transformation,
\begin{align}
b_j = \frac{1}{\sqrt{N_{\rm FI}}} \sum_{\bm k} e^{i{\bm k}\cdot {\bm r}_j} b_{\bm k},
\end{align}
can be used to approximate the Hamiltonian of the FI as
\begin{align}
H_{\rm FI} &= \sum_{\bm{q}}\hbar\omega_{\bm{q}}b^{\dagger}_{\bm{q}}b_{\bm{q}},\\
\hbar\omega_{\bm{q}} &= \mathcal{D}\bm{q}^2+\hbar|\gamma_{\rm g}| h_{\rm dc},
\end{align}
where $N_{\rm FI}$ is the number of unit cells in the FI, $\hbar\omega_{\bm{q}}$ is energy dispersion of a magnon, and $\mathcal{D}$ is spin stiffness. 
Since FMR is used to excite uniform spin precession by microwave irradiation, the Hamiltonian of the FI can be approximated as
\begin{align}
H_{\rm FI} &=\hbar\omega_{\bm{0}}b^{\dagger}_{\bm{0}}b_{\bm{0}},
\end{align}
where $\omega_{\bm{0}} =|\gamma_{\rm g}| h_{\rm dc}$($>0$).

\subsection{FI/2DEG Interface}
\label{sec:interface}

Spin operators for the conduction electrons in the laboratory coordinates are first defined as
\begin{align}
s^a_{\bm{q}}&= \sum_{\sigma,\sigma'}\sum_{\bm{k}}c^{\dagger}_{\bm{k}\sigma}(\hat{\sigma}_a)_{\sigma\sigma'}c_{\bm{k}+\bm{q}\sigma}, \quad (a= x, y, z),
\end{align}
where $\hat{\sigma}_a$ ($a=x,y,z$) are the Pauli matrices. In addition, spin operators in the coordinates fixed to the direction of the ordered spin are defined as
\begin{align}
\begin{pmatrix}
s^{x'}_i \\ s^{y'}_i \\ s^{z'}_i
\end{pmatrix}
&=
\begin{pmatrix}
\cos\theta & \sin\theta & 0
\\ -\sin\theta & \cos\theta & 0 
\\ 0 & 0 & 1
\end{pmatrix}
\begin{pmatrix}
s^{x}_i \\ s^{y}_i \\ s^{z}_i
\end{pmatrix} .
\end{align}
In the new coordinates $(x',y',z')$, the spin ladder operators are defined as
\begin{align}
s^{x'\pm}_{\bm{q}}
&=\frac{1}{2}\sum_{\sigma,\sigma'}\sum_{\bm{k}}c^{\dagger}_{\bm{k}\sigma}(\hat{\sigma}^{x'\pm})_{\sigma\sigma'}c_{\bm{k}\pm\bm{q}\sigma'},\\
\hat{\sigma}^{x'\pm}& = \hat{\sigma}^{y'} \pm i \hat{\sigma}^{z'} \nonumber \\
&=-\sin\theta~\hat{\sigma}_x +\cos\theta~\hat{\sigma}_y \pm i\hat{\sigma}_z , \label{eq:hatsig}
\end{align}
and used to write the Hamiltonian of the interfacial exchange coupling between the FI and 2DEG as \cite{Ohnuma2014, Matsuo2018, Kato2019, Kato2020, Ominato2020a, Ominato2020b, Ominato2022, Funato2022, Tajima2022, Yama2021, Yama2023}
\begin{align}
H_{\rm int}= \sum_{\bm{q}}(\mathcal{T}_{\bm{q}}S^{x'+}_{\bm{q}}s^{x'-}_{\bm{q}}+\mathcal{T}^*_{\bm{q}}s^{x'+}_{\bm{q}} S^{x'-}_{\bm{q}}), \label{eq:Hint1}
\end{align}
where $\mathcal{T}_{\bm{q}}$ represents the magnitude of exchange interaction~\footnote{The $x'$ component of the interfacial exchange coupling is dropped because this term, which is approximated as $\sum_{\bm q} 2\mathcal{T}_{\bm q} s_{\bm q}^{x'} \langle S_{\bm q}^{x'} \rangle \simeq \sum_{\bm q} 2S_0 \mathcal{T}_{\bm q} s_{\bm q}^{x'}$, works as an effective Zeeman field on the conduction electrons. It is assumed that this effective field, which is called exchange bias, is much smaller than other energy scales such as temperature and spin-orbit interactions.}. For a uniform spin precession of the FI, the Hamiltonian of the interface with the uniform contribution ($\bm{q}=0$) is approximated as
\begin{align}
H_{\rm int} &= \mathcal{T}_{\bm{0}}S^{x'+}_{\bm{0}}s^{x'-}_{\bm{0}}+\mathcal{T}^*_{\bm{0}}s^{x'+}_{\bm{0}} S^{x'-}_{\bm{0}}\nonumber \\
&= \sqrt{2S_0}(\mathcal{T}_{\bm{0}}b_{\bm{0}} s^{x'-}_{\bm{0}}+\mathcal{T}^*_{\bm{0}}s^{x'+}_{\bm{0}}b^\dagger_{\bm{0}}). \label{eq:Hint2}
\end{align}

\section{Formulation}
\label{sec:formulation}

Spin-charge conversion in the 2DEG is formulated by using the Boltzmann equation in this section. First, the Boltzmann equation is introduced, and the assumptions used in our calculation in Sec.~\ref{sec:Boltzmann} are explained. Next, explicit forms of the two collision terms in Sec.~\ref{sec:spinpumping} and Sec.~\ref{sec:impurity} are derived. Finally, the Boltzmann equation is solved, and the spin and current densities induced by the IREE are derived in Sec.~\ref{sec:MandI}.

\subsection{Boltzmann equation}
\label{sec:Boltzmann}

In the present model, the distribution function in the Boltzmann equation becomes a matrix in general, reflecting spin polarization caused by both the effective Zeeman field and external driving. The formulation is simplified by assuming that the spin-orbit interaction is much larger than damping rate $\Gamma$. Under this assumption, the distribution function can be approximated as a diagonal matrix in the eigenstate basis $|{\bm k},\gamma\rangle$ introduced for the conduction electrons in Sec.~\ref{sec:2DEG}, and it is denoted as $f({\bm k},\gamma)$ in uniform steady state~\cite{Suzuki2023}. The Boltzmann equation for our model is then described as
\begin{align}
0 = \frac{\partial f}{\partial t}\biggl{|}_{\rm pump}+\frac{\partial f}{\partial t}\biggl{|}_{\rm imp} , \label{eq:0ecoll}
\end{align}
where $\partial f/\partial t|_{\rm pump}$ is a collision term due to spin injection from the FI into the 2DEG through the interface, and $\partial f/\partial t|_{\rm imp}$ is a collision term due to impurity scattering. Note that this formulation does not require spin current whose definition is subtle\cite{Khaetskii2006, Shitade2022} for electron systems with spin-orbit interactions.

For the linear response to the external driving, it is sufficient to consider a non-equilibrium distribution function with an energy shift depending on wave-number vector $\bm{k}$ and spin $\gamma$ as \cite{Wilson1953, Ziman1960, Lundstrom2000}
\begin{align}
f(\bm{k},\gamma)=f_0(\epsilon_{\bm{k}}+\gamma h_{\rm eff}(\varphi)-\mu-\delta\mu(\bm{k},\gamma)),
\end{align}
where $f_0(\epsilon) =(\exp[\tempinv (\epsilon-\mu)]+1)^{-1}$ is the Fermi distribution function, and $\tempinv$ is inverse temperature. Energy shift $\delta \mu(\bm{k},\gamma)$ can be regarded as a nonequilibrium chemical potential driven by spin pumping. Hereafter, the linear response of the 2DEG with respect to spin pumping is investigated. For this investigation, it is sufficient to approximate the distribution function as
\begin{align}
f(\bm{k},\gamma) \simeq f_0(E^{\gamma}_{\bm{k}})-\frac{\partial f_0(E^{\gamma}_{\bm{k}})}{\partial E^{\gamma}_{\bm{k}}} \delta\mu(\bm{k},\gamma).
\label{eq:fexpan}
\end{align}

\subsection{Collision term due to spin pumping}
\label{sec:spinpumping}

Spin injection from the interface into the 2DEG is described by stochastic excitation induced by magnon absorption and emission. This process can be expressed by the collision term as
\begin{align}
& \frac{\partial f(\bm{k},\gamma)}{\partial t}\biggl{|}_{\rm pump} \nonumber \\
&=\sum_{\bm{k}'}\sum_{\gamma'=\pm}
\Bigl{[}
P_{\bm{k}'\gamma'\rightarrow\bm{k}\gamma}
f(\bm{k}',\gamma')(1-f(\bm{k},\gamma))\nonumber \\
&\hspace{15mm}
-P_{\bm{k}\gamma\rightarrow\bm{k}'\gamma'}
f(\bm{k},\gamma)(1-f(\bm{k}',\gamma'))
\Bigl{]}, \label{eq:colpump1}
\end{align}
where $P_{\bm{k}\gamma\rightarrow\bm{k}'\gamma'}$ is the transition rate from initial state $|\bm{k},\gamma\rangle$ to final state $|\bm{k}',\gamma'\rangle$. Transition rate is calculated with Fermi’s golden rule as
\begin{align}
&P_{\bm{k}\gamma\rightarrow\bm{k}'\gamma'} \nonumber \\
&=\sum_{N_{\bm{0}}}\sum_{\Delta N_{\bm{0}} = \pm 1}
\frac{2\pi}{\hbar}
\Bigl{|}\langle \bm{k}',\gamma'|\langle N_{\bm{0}}+\Delta N_{\bm{0}} |H_{\rm int}
|\bm{k},\gamma\rangle|N_{\bm{0}}\rangle \Bigl{|}^2\nonumber \\
&\hspace{5mm}\times \rho(N_{\bm{0}}) \delta\Bigl{(}E^{\gamma'}_{\bm{k}'}-E^{\gamma}_{\bm{k}}+\Delta N_{\bm{0}}\hbar\omega_{\bm{0}}\Bigl{)}, \label{eq:Pkkp}
\end{align}
where $|N_{\bm{0}}\rangle$ is the eigenstate of the magnon number operator, i.e., $b^{\dagger}_{\bm{0}}b_{\bm{0}}|N_{\bm{0}}\rangle=N_{\bm{0}}|N_{\bm{0}}\rangle$, $\Delta N_{\bm{0}}=\pm 1$ is a change of the magnon number, and $\rho(N_{\bm{0}})$ describes a nonequilibrium distribution function for the uniform spin precession induced by external microwaves. It is assumed that distribution function $\rho(N_{\bm{0}})$ has a sharp peak at its average $\langle N_{\bm{0}}\rangle$ and $\langle N_{\bm{0}}\rangle \gg 1$. 
We note that for the Hamiltonian $H_{\rm int}$ given in Eq.~(\ref{eq:Hint2}), the transition rate $P_{{\bm k}\gamma\rightarrow {\bm k}'\gamma'}$ is nonzero only for $\bm{k}'={\bm k}$.
The summation can then be approximated as
\begin{align}
\sum_{N_{\bm{0}}} \rho(N_{\bm{0}}) F(N_{\bm{0}})
\simeq F(\langle N_{\bm{0}}\rangle),
\label{eq:appN}
\end{align}
where $F(x)$ is an arbitrary function. In this approximation, since the transition rate is proportional to $\langle N_{\bm{0}}\rangle$, $\langle N_{\bm{0}}\rangle$ represents the strength of spin pumping.

In Sec.~\ref{sec:MandI}, the current induced by spin pumping is evaluated up to the linear response with respect to $\langle N_{\bm{0}}\rangle$. For this evaluation, it is sufficient to evaluate $\delta\mu(\bm{k},\gamma)$ up to the linear contribution of $\langle N_{\bm{0}}\rangle$~\footnote{Note that the feedback from the spin injection into the spin-precession states can be neglected since it only affects the higher-order contributions with respect to $\langle N_{\bm{0}}\rangle$.}. 
In the following calculation, it is essential that the transition rate depends on the overlap of the spinor wavefunctions between the initial and final states, $\langle {\bm k}',\gamma'|{\bm k},\gamma\rangle$, which is written in terms of the coefficients $C_{\sigma\gamma}$ given in Eq.~(\ref{eq:Cupdown}). Note that this formulation is valid only when the spin-orbit interaction is much larger than the energy broadening due to impurity scattering, i.e., ${\rm max} \, (k_{\rm F}\alpha, k_{\rm F}\beta) \gg \Gamma$.

By straightforward calculation, the collision term due to spin pumping is obtained as
\begin{align}
&\frac{\partial f(\bm{k},\gamma)}{\partial t}\biggl{|}_{\rm pump}
=-\frac{\pi S_0|\mathcal{T}_{\bm{0}}|^2\langle N_{\bm{0}}\rangle\gamma}{\hbar}
\sum_{\gamma'=\pm}\nonumber \\
&\times
\gamma'\biggl{[}(\hat{\bm{h}}_{\rm eff}(\varphi)\cdot\hat{\bm{m}}(\theta)-\gamma')\cdot(\hat{\bm{h}}_{\rm eff}(\varphi)\cdot\hat{\bm{m}}(\theta)+\gamma)\nonumber \\
&\hspace{8mm}\times
[f_0(E^{\gamma}_{\bm{k}}-\hbar\omega_{\bm{0}})-f_0(E^{\gamma}_{\bm{k}})]
\delta((\gamma'-\gamma)h_{\rm eff}(\varphi)+\hbar\omega_{\bm{0}})
\nonumber \\
&\hspace{5mm}
+(\hat{\bm{h}}_{\rm eff}(\varphi)\cdot\hat{\bm{m}}(\theta)+\gamma')\cdot(\hat{\bm{h}}_{\rm eff}(\varphi)\cdot\hat{\bm{m}}(\theta)-\gamma)\nonumber \\
&\hspace{8mm}\times
[f_0(E^{\gamma}_{\bm{k}}+\hbar\omega_{\bm{0}})-f_0(E^{\gamma}_{\bm{k}})]
\delta((\gamma'-\gamma)h_{\rm eff}(\varphi)-\hbar\omega_{\bm{0}})
\biggl{]}, \label{eq:colpumfin}
\end{align}
where $\hat{\bm{h}}_{\rm eff}(\varphi)\equiv \bm{h}_{\rm eff}(\varphi)/h_{\rm eff}(\varphi)$ is the direction of the effective Zeeman field imposed on the 2DEG electrons, and $\hat{\bm{m}}(\theta)=(\cos\theta,\sin\theta)^T$ is the direction of the localized spin in the FI. Note that to use Fermi's golden rule in Eq. (\ref{eq:Pkkp}), in addition to assuming the weak-scattering condition, it is assumed that $\mathcal{T}_{\bm{0}}$ is much smaller than the Rashba and Dresselhaus spin-orbit interactions. For detailed derivation, see Appendices~\ref{app:CollisionTerms} and \ref{app:off-diagonal}.

\subsection{Collision term due to impurity scattering}
\label{sec:impurity}

The collision term due to impurity scattering is written as
\begin{align}
& \frac{\partial f(\bm{k},\gamma)}{\partial t}\biggl{|}_{\rm imp} \nonumber \\
&=\sum_{\bm{k}'}\sum_{\gamma'=\pm}
\Bigl{[}
Q_{\bm{k}'\gamma'\rightarrow\bm{k}\gamma}
f(\bm{k}',\gamma')(1-f(\bm{k},\gamma))\nonumber \\
&\hspace{15mm}
-Q_{\bm{k}\gamma\rightarrow\bm{k}'\gamma'}
f(\bm{k},\gamma)(1-f(\bm{k}',\gamma'))
\Bigl{]}, \label{eq:colimp1}
\end{align}
where $Q_{\bm{k}\gamma\rightarrow\bm{k}'\gamma'}$ is the transition rate of electron scattering from initial state $|\bm{k},\gamma\rangle$ to final state $|\bm{k}',\gamma'\rangle$. 
According to the Born approximation (or Fermi's golden rule), the transition rate is given as
\begin{align}
&Q_{\bm{k}\gamma\rightarrow\bm{k}'\gamma'} \nonumber \\
&= \frac{2\pi}{\hbar}
\Bigl{|}\langle \bm{k}',\gamma'|H_{\rm imp}({\bm R})|\bm{k},\gamma\rangle \Bigl{|}^2 \delta\Bigl{(}E^{\gamma'}_{\bm{k}'}-E^{\gamma}_{\bm{k}}\Bigl{)}. \label{eq:Qkkp}
\end{align}
Note that the transition rate due to impurity scattering also includes the overlap of the spin states between the initial and final states.

The collision term due to impurity scattering is calculated as
\begin{align}
&\frac{\partial f(\bm{k},\gamma)}{\partial t}\biggl{|}_{\rm imp}
\simeq
\frac{\Gamma}{2\hbar k_{\rm F}}
\sum_{\gamma'=\pm}\int_0^{2\pi}\frac{d\varphi'}{2\pi}k(\varphi',\gamma')\nonumber \\
&\hspace{18mm}\times
[1+\gamma\gamma'\hat{\bm{h}}_{\rm eff}(\varphi)\cdot\hat{\bm{h}}_{\rm eff}(\varphi')]\nonumber \\
&\hspace{18mm}\times
[\delta\mu(\varphi',\gamma')-\delta\mu(\bm{k},\gamma)]\delta(E^{\gamma}_{\bm{k}}-\mu),
\label{eq:goldenimp}
\end{align}
where $\Gamma = 2\pi n_{\rm imp} u^2 D(\epsilon_{\rm F})$ is level broadening due to the impurities, $n_{\rm imp}$ is impurity density, $D(\epsilon_{\rm F})=m^*/(2\pi \hbar^2)=k_{\rm F}/(2\pi\hbar v_{\rm F})$ is the density of states per unit area, and $v_{\rm F}=\hbar k_{\rm F}/m^*$ is the Fermi velocity. 
Here, the Fermi wavenumber of electrons with azimuth angle $\varphi$ and spin $\gamma$ in the absence of microwave driving and the corresponding chemical potential shift are defined as
\begin{align}
k(\varphi,\gamma) &= k_{\rm F}-2\pi \gamma D(\epsilon_{\rm F})\sqrt{\alpha^2+\beta^2+2\alpha\beta\sin2\varphi}, \\
\delta\mu(\varphi,\gamma) &= \delta\mu(|\bm{k}|,\gamma)|_{|\bm{k}|=k(\varphi,\gamma)},
\end{align}
respectively. In the derivation of Eq.~(\ref{eq:goldenimp}), the sum over the wavenumber is replaced with the integral as
\begin{align}
\frac{1}{\mathcal{A}}\sum_{\bm{k}} (\cdots)  = \frac{1}{2\pi}\int_{0}^{\infty}dk~|\bm{k}|\int_{0}^{2\pi}\frac{d\varphi} {2\pi}(\cdots), \label{eq:sumint}
\end{align}
and 
\begin{align}
-\frac{\partial f_0(E^\gamma_{\bm{k}})}{\partial E^\gamma_{\bm{k}}} &\simeq \delta(E^\gamma_{\bm{k}}-\mu) \simeq \frac{1}{\hbar v_{\rm F}}\delta(|\bm{k}|-k(\varphi,\gamma)).
\end{align}
was used. For a detailed derivation, see Appendices~\ref{app:CollisionTerms} and \ref{app:off-diagonal}.

\subsection{Current in 2DEG induced by IREE}
\label{sec:MandI}

Integrating Eq.~(\ref{eq:0ecoll}) over $\epsilon_{\bm k}$ with the two collision terms presented in the previous two subsections gives the equation for $\delta\mu(\varphi,\gamma)$ as
\begin{align}
&\delta\mu(\varphi,\gamma)
=\frac{k_{\rm F}}{k(\varphi,\gamma)}\gamma G(\varphi,\gamma,\theta,\hbar\omega_{\bm{0}})\nonumber \\
&+\frac{1}{2}\sum_{\gamma'=\pm}
\int_{0}^{2\pi}\frac{d\varphi'}{2\pi}\frac{k(\varphi',\gamma')}{k_{\rm F}}[1+\gamma\gamma'\hat{\bm{h}}_{\rm eff}(\varphi)\cdot\hat{\bm{h}}_{\rm eff}(\varphi')]\nonumber\\
&\hspace{5mm}\times\delta\mu(\varphi',\gamma'),\label{eq:Phieq} 
\end{align}
where 
\begin{align}
&G(\varphi,\gamma,\theta,\hbar\omega_{\bm{0}})
\equiv-\frac{\pi S_0|\mathcal{T}_{\bm{0}}|^2 \langle N_{\bm{0}}\rangle \hbar\omega_{\bm{0}}}{\Gamma}
\sum_{\gamma'=\pm}\nonumber \\
&\times
\gamma'\Bigl{[}(\hat{\bm{h}}_{\rm eff}(\varphi)\cdot\hat{\bm{m}}(\theta)-\gamma')\cdot(\hat{\bm{h}}_{\rm eff}(\varphi)\cdot\hat{\bm{m}}(\theta)+\gamma)\cdot L^+\nonumber \\
&\hspace{4mm}
-(\hat{\bm{h}}_{\rm eff}(\varphi)\cdot\hat{\bm{m}}(\theta)+\gamma')\cdot(\hat{\bm{h}}_{\rm eff}(\varphi)\cdot\hat{\bm{m}}(\theta)-\gamma)\cdot L^-\Bigl{]},\label{eq:defG}
\end{align}
represents the effect of spin pumping. 
If finite energy broadening due to impurity scattering is taken into account, the $\delta$ function in the transition rate can be replaced with the spectral functions as follows\cite{Yama2021, Yama2023}:
\begin{align}
L^{\pm} = \frac{\Gamma/2\pi}{(\hbar\omega_{\bm{0}}\pm(\gamma'-\gamma)h_{\rm eff}(\varphi))^2+(\Gamma/2)^2}.
\end{align}

The solution of Eq.~(\ref{eq:Phieq}) is given as
\begin{align}
&\delta\mu(\varphi,\gamma)=
\frac{k_{\rm F}}{k(\varphi,\gamma)}\gamma G(\varphi,\gamma,\theta,\hbar\omega_{\bm{0}})\nonumber \\
&+\frac{\gamma}{2}\hat{\bm{h}}^T_{\rm eff}(\varphi)\biggl{(}\hat{I}-\int_{0}^{2\pi}\frac{d\varphi'}{2\pi}\hat{\bm{h}}_{\rm eff}(\varphi')\cdot \hat{\bm{h}}_{\rm eff}^{T}(\varphi')\biggl{)}^{-1}\nonumber \\
&\hspace{5mm}\times\int_{0}^{2\pi}\frac{d\varphi''}{2\pi}\sum_{\gamma''=\pm}\hat{\bm{h}}_{\rm eff}(\varphi'')G(\varphi'',\gamma'',\theta,\hbar\omega_{\bm{0}}), \label{eq:soldelmu}
\end{align}
where $\hat{I}$ is an identity matrix, $\bm{a} \cdot \bm{a}^T$ indicates 
\begin{align}
\bm{a} \cdot \bm{a}^T = \left( \begin{array}{c} a_x \\ a_y \end{array} \right) (a_x \ a_y) 
= \left( \begin{array}{cc} a_x a_x & a_x a_y \\ a_y a_x & a_y a_y \end{array} \right) ,
\end{align}
and $\hat{A}^{-1}$ indicates the inverse matrix of $\hat{A}$. Note that it is assumed that $\delta\mu(\varphi,\gamma)$ of Eq. (\ref{eq:soldelmu}) does not change the number of electrons in the 2DEG; i.e., 
\begin{align}
0=\sum_{\bm{k}}\sum_{\gamma=\pm}\delta\mu(\bm{k},\gamma)\delta(E^\gamma_{\bm{k}}-\mu).
\end{align}
is satisfied. This solution for $\delta\mu(\varphi,\gamma)$ can be used to calculate the spin and current densities in the 2DEG induced by spin pumping as follows. The spin density in the 2DEG is expressed up to the linear order of $\delta\mu(\varphi,\gamma)$ as
\begin{align}
&\bm{s}=
\frac{\hbar}{2\mathcal{A}}\sum_{\bm{k},\gamma}\langle \bm{k},\gamma|\bm{\sigma}|\bm{k},\gamma\rangle f(\bm{k},\gamma)\nonumber \\
&=-\frac{\hbar D(\epsilon_{\rm F})}{2}\sum_{\gamma=\pm}\int_0^{2\pi}\frac{d\varphi}{2\pi}\frac{k(\varphi,\gamma)}{k_{\rm F}}\delta\mu(\varphi,\gamma)\gamma\hat{\bm{h}}_{\rm eff}(\varphi).
\label{eq:MIREE}
\end{align}
In a similar way, the current density induced in the 2DEG is expressed as 
\begin{align}
&\bm{j}=\frac{e}{\mathcal{A}} \sum_{\bm{k},\gamma}\bm{v}(\bm{k},\gamma)f(\bm{k},\gamma)\nonumber \\
&=eD(\epsilon_{\rm F})\sum_{\gamma=\pm}\int_0^{2\pi}\frac{d\varphi}{2\pi}\frac{k(\varphi,\gamma)}{k_{\rm F}}\delta\mu(\varphi,\gamma)\bm{v}(\bm{k},\gamma)|_{|\bm{k}|=k(\varphi,\gamma)},
\label{eq:IIREE}
\end{align}
where $e$ ($<0$) is electron charge and $\bm{v}(\bm{k},\gamma)$ is electron velocity defined as 
\begin{align}
\bm{v}(\bm{k},\gamma)&=
\frac{1}{\hbar}\frac{\partial E^{\gamma}_{\bm{k}}}{\partial\bm{k}}
=\frac{\hbar\bm{k}}{m^*}+\frac{\gamma}{\hbar}\frac{\partial h_{\rm eff}(\bm{k})}{\partial\bm{k}} .
\end{align}
Note that the current density induced by IREE is formulated without using spin current.

\section{Results}
\label{sec:results}

In this section, spin and current densities in the 2DEG induced by spin pumping are calculated for the four specific cases. Dependence of the maximum values of the spin and current densities on $\alpha/\beta$ is also discussed in Sec.~\ref{sec:MIamp}. In the following, spin density is expressed in units of $s_\alpha\equiv \pi\hbar D(\epsilon_{\rm F})S_0|\mathcal{T}_{\bm{0}}|^2 \langle N_{\bm{0}}\rangle/(2k_{\rm F}\alpha)$ or $s_\beta\equiv \pi\hbar D(\epsilon_{\rm F})S_0|\mathcal{T}_{\bm{0}}|^2 \langle N_{\bm{0}}\rangle/(2k_{\rm F}\beta)$, and current density is expressed in units of $j_0=\pi|e|D(\epsilon_{\rm F})S_0|\mathcal{T}_{\bm{0}}|^2 \langle N_{\bm{0}}\rangle/(\hbar k_{\rm F})$.

\subsection{Rashba spin-orbit interaction ($\alpha/\beta=\infty$)}
\label{sec:Rashba}

\begin{figure*}[tb]
\centering
\includegraphics[width=145mm]{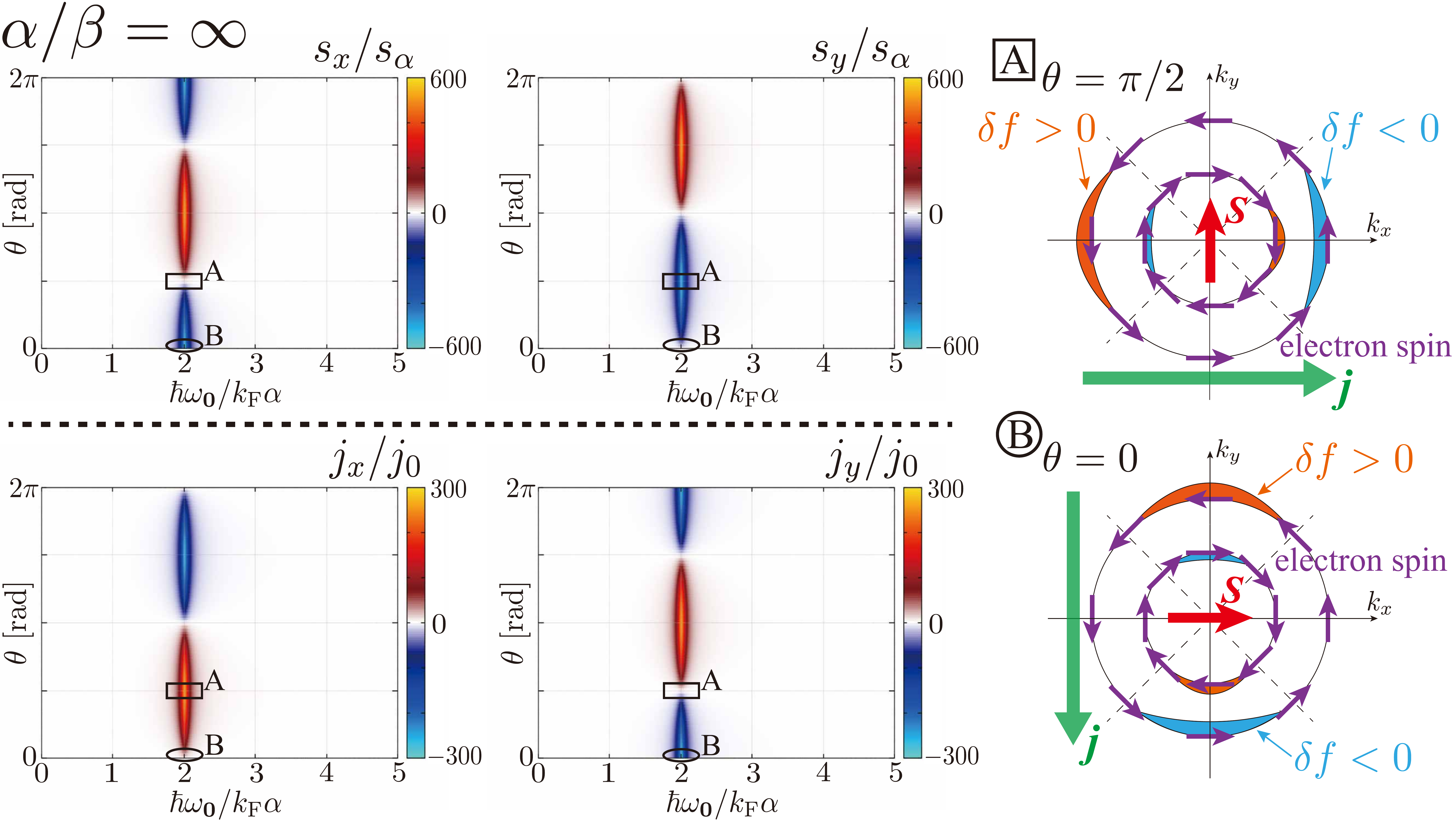}
\caption{Color plots of spin density $\bm{s}=(s_x, s_y)$ (two upper panels) and current density $\bm{j}=(j_x, j_y)$ (two lower panels) as a function of FMR frequency $\omega_{\bm0}$ and azimuth angle $\theta$ of spin polarization in the FI for the Rashba spin-orbit interaction ($\alpha/\beta=\infty$). The two right insets schematically show the change of the distribution function and the direction of the current for $\theta = \pi/2$ and $\theta = 0$, respectively. 
Note that $\delta f$ represents a change of the distribution function of the 2DEG electrons, $\delta f\equiv f(\bm{k},\gamma)-f_0(E^{\gamma}_{\bm{k}})$.
It was assumed that $\Gamma/k_{\rm F}\alpha$=0.1.} 
\label{fig:Rashba}
\end{figure*}

The case that only Rashba spin-orbit interaction exists ($\beta = 0$) is discussed first. In this case, the effective Zeeman field $h_{\rm eff}(\varphi) = k_{\rm F}\alpha$ becomes independent of $\varphi$. The four color plots in Fig.~\ref{fig:Rashba} show spin density $\bm{s}/s_\alpha=(s_x/s_\alpha,s_y/s_\alpha)$ and current density $\bm{j}/j_{0}=(j_x/j_{0},j_y/j_{0})$ in the 2DEG as functions of FMR frequency $\omega_{\bm{0}}$ and the azimuth, $\theta$, of the spin in the FI. Both the spin and current densities peak at $\hbar\omega_{\bm{0}}=2k_{\rm F}\alpha$ ($=2h_{\rm eff})$, i.e., when spin-splitting energy matches microwave energy.

The peak height of the spin and current densities depends on $\theta$. For $\theta=\pi/2$ (indicated by A in a square in each color plot), spin density ${\bm s}$ is induced in the $-y$ direction, while current density ${\bm j}$ is induced in the $+x$ direction. This result can be explained intuitively as follows. The spin in the $-y$ direction is injected from the FI into the 2DEG when $\theta=\pi/2$. That injection of spin induces spin density ${\bm s}$ in the $-y$ direction. Note that the direction of spins injected from the FI into the 2DEG is opposite to that of the localized spin in the FI, $\bm{S}$, since spin transfer is induced by spin relaxation in the FI.
As a result, the nonequilibrium distribution function increases (decreases) the $-y$-spin ($+y$-spin) band. The region of the Fermi surface on which the distribution function increases (decreases) is shown schematically with the orange (blue) in the upper-right inset in Fig.~\ref{fig:Rashba}. Since the density of states of the outer Fermi surface is larger than that of the inner one, this change of the distribution function produces a net flow of electrons in the $-x$ direction, which produces a current in the $+x$ direction. For $\theta=0$ (indicated by the ellipse B in each color plot), spin density ${\bm s}$ is induced in the $-x$ direction, while current density ${\bm j}$ flows in the $-y$ direction. This result is also explained intuitively in the same way as for $\theta = \pi/2$ (see the lower-right inset in Fig.~\ref{fig:Rashba}).

The phenomenon obtained here can be regarded as IREE due to  spin density in the 2DEG, which is induced by spin pumping from the FI, i.e., spin injection through electron-spin flipping at the interface. While the concept of the spin current may be helpful for intuitive understanding of this phenomenon, it is remarkable that the induced current is calculated without introducing it in our study.

\subsection{Dresselhaus spin-orbit interaction ($\alpha/\beta=0$)}
\label{sec:Dresselhaus}

\begin{figure*}[tb]
\centering
\includegraphics[width=145mm]{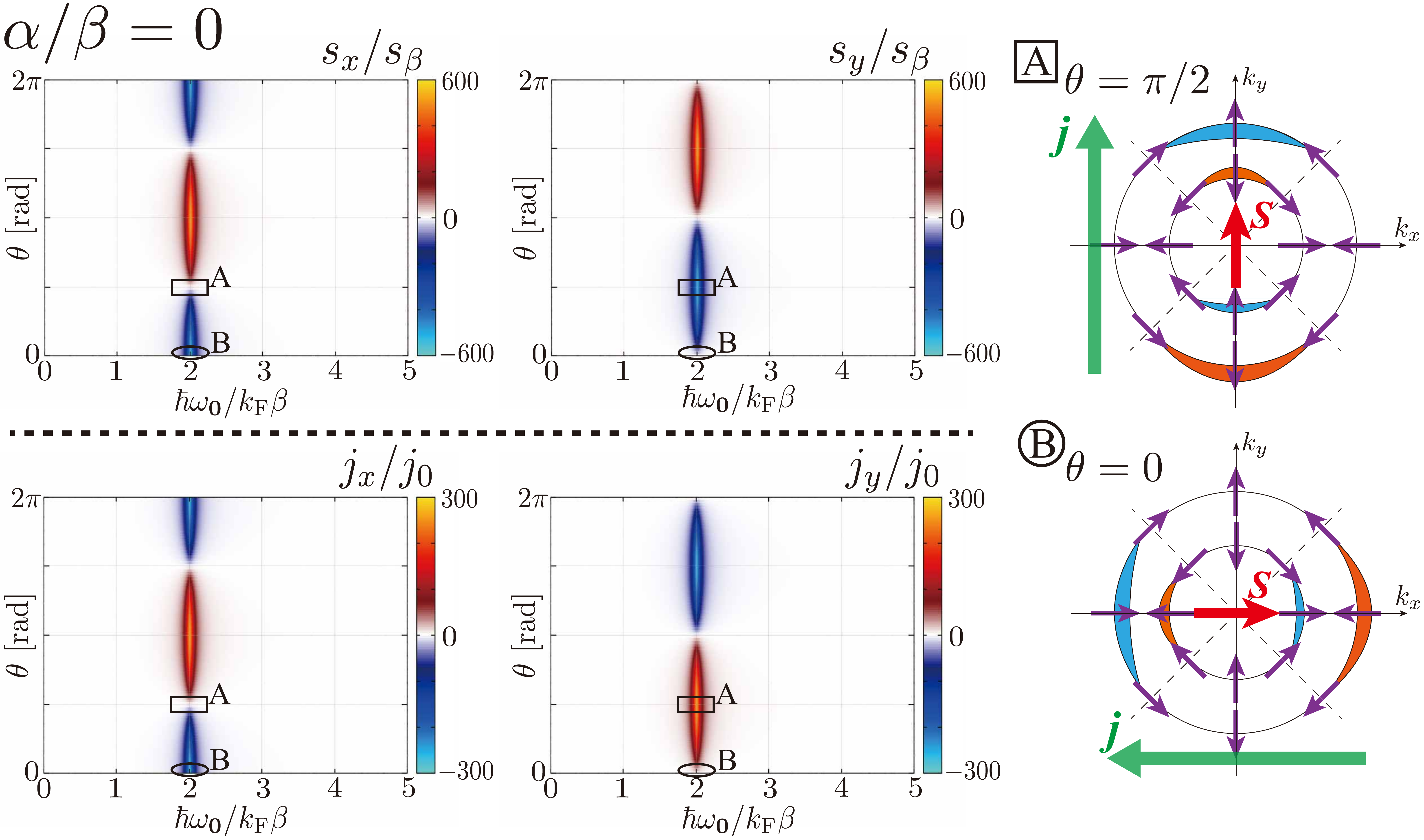}
\caption{Color plots of spin density $\bm{s}=(s_x, s_y)$ (two upper panels) and current density $\bm{j}=(j_x, j_y)$ (two lower panels) for the Dresselhaus spin-orbit interaction ($\alpha/\beta=0$). The two right 
insets schematically show the change of the distribution function and the direction of the current for $\theta = \pi/2$ and $\theta = 0$, respectively. It was assumed that $\Gamma/k_{\rm F}\beta$=0.1.}
\label{fig:Dresselhaus}
\end{figure*}

The case that only the Dresselhaus spin-orbit interaction exists ($\alpha = 0$) is discussed next. Also in this case, effective Zeeman field $h_{\rm eff}(\varphi) = k_{\rm F}\beta$ becomes independent of $\varphi$. The four color plots in Fig.~\ref{fig:Dresselhaus} show spin density $\bm{s}/s_\beta=(s_x/s_\beta,s_y/s_\beta)$ and current density $\bm{j}/j_{0}=(j_x/j_{0},j_y/j_{0})$ as functions of FMR frequency $\omega_{\bm{0}}$ and spin azimuth $\theta$. Both the spin and current densities peak at $\hbar\omega_{\bm{0}}=2k_{\rm F}\beta$ ($=2h_{\rm eff})$ as in the previous case of Rashba spin-orbit interaction ($\beta = 0$). On the contrary, the dependence of $\bm{j}$ on $\theta$ differs from that in the previous case. For $\theta=\pi/2$ (indicated by  A in a square), spin density ${\bm s}$ is induced in the $-y$ direction, while current density ${\bm j}$ is induced in the $+y$ direction. This result can be explained intuitively in the same way as the previous case as follows. The spin in the $-y$ direction is injected from the FI into the 2DEG, and that spin injection induces spin density ${\bm{s}}$ in the $-y$ direction. The change of the nonequilibrium distribution function (see upper-right inset in Fig.~\ref{fig:Dresselhaus}) produces a net flow of electrons in the $-y$ direction, which generates a current in the $+y$ direction. For $\theta=0$ (indicated by the ellipse B), spin density ${\bm s}$ is induced in the $-x$ direction, while current density ${\bm j}$ flows in the $-x$ direction. 
This is explained schematically in the lower-right inset in Fig.~\ref{fig:Dresselhaus}.

Notably, $\bm{j}$ is orthogonal to ${\bm s}$ in the case of Rashba spin-orbit interaction, while it is parallel to ${\bm s}$ in the case of Dresselhaus spin-orbit interaction. These contrasting behaviors are clearly due to the difference in the spin texture on the Fermi surface, which is determined by the effective Zeeman field ${\bm h}_{\rm eff}$ (as shown by comparing the right insets in Figs.~\ref{fig:Rashba} and \ref{fig:Dresselhaus}).

\subsection{Case of $\alpha/\beta=1.1$}
\label{sec:aob11}

\begin{figure*}[tb]
\centering
\includegraphics[width=145mm]{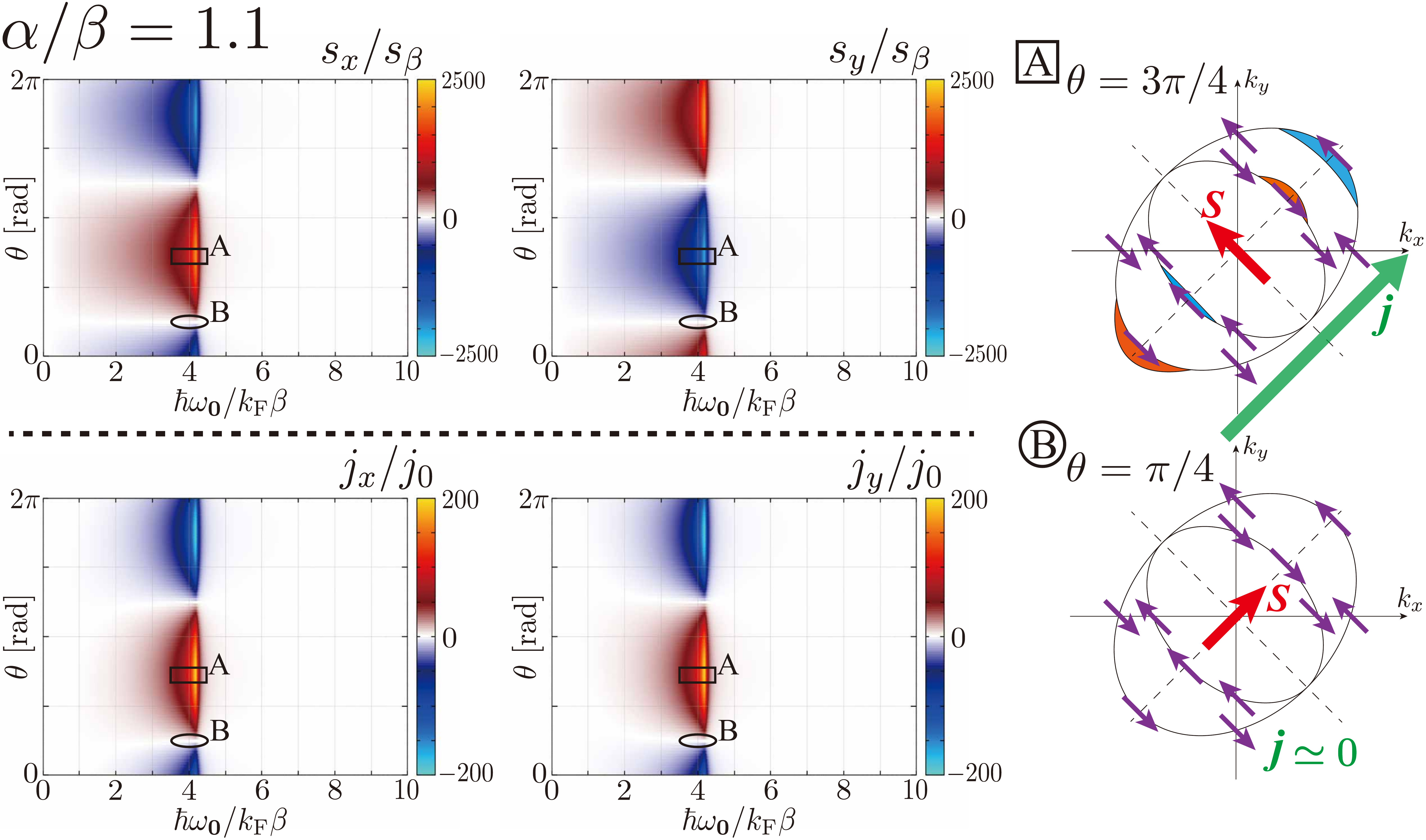}
\caption{Color plots of spin density $\bm{s}=(s_x, s_y)$ (two upper panels) and the current density $\bm{j}=(j_x, j_y)$ (two lower panels) in the case of $\alpha/\beta=1.1$. The two right insets schematically show the change of the distribution function and the direction of the current for $\theta = 3\pi/4$ and $\theta = \pi/4$, respectively. It was assumed that $\Gamma/k_{\rm F}\beta=0.1$.}
\label{fig:aob11}
\end{figure*}

When the Rashba and Dresselhaus spin-orbit interactions 
compete ($\alpha \simeq \beta$), the spin and current densities induced by spin pumping change significantly. To investigate this difference, the case of $\alpha/\beta=1.1$ is considered as follows. Effective Zeeman field $h_{\rm eff}(\varphi)$ depends on $\varphi$ and changes in the range of $0 \lesssim h_{\rm eff}(\varphi) \lesssim 2 k_{\rm F}\beta$. The four color plots in Fig.~\ref{fig:aob11} show spin density $\bm{s}/s_\beta=(s_x/s_\beta,s_y/s_\beta)$ and current density $\bm{j}/j_{0}=(j_x/j_{0},j_y/j_{0})$ as functions of $\omega_{\bm{0}}$ and $\theta$. Corresponding to the distribution of $h_{\rm eff}(\varphi)$, both the spin and current densities are induced in a wide range of $0\lesssim \hbar\omega_{\bm{0}}\lesssim 4 k_{\rm F}\beta$, and their amplitudes take maxima at $\hbar\omega_{\bm{0}}\simeq 4k_{\rm F}\beta$. Spin density ${\bm s}$ always points in the $(1,-1)$ direction, while current $\bm j$ flows in the $(1,1)$ direction.

The amplitudes of ${\bm s}$ and ${\bm j}$ depend on spin azimuth $\theta$: they take maxima at $\theta=3\pi/4$ or $7\pi/4$ and almost vanish at $\theta=\pi/4$ or $5\pi/4$. This distinctive result can be explained as follows (see upper and lower-right insets in Fig.~\ref{fig:aob11}). For $\theta =3\pi/4$ (indicated by A in a square in Fig.~\ref{fig:aob11}), the $7\pi/4$ component of spin density increases, resulting in a change of the distribution function in the direction of $\varphi = \pi/4$ and $5\pi/4$. This change in the distribution function causes a net electron flow (a current) in the direction of $\varphi = 5\pi/4$ ($\varphi = \pi/4$). On the contrary, for $\theta =\pi/4$ (indicated by B in a circle in Fig.~\ref{fig:aob11}), the spin in the direction of $\pi/4$ cannot enter the 2DEG because it is always perpendicular to the effective Zeeman field, i.e., the spin polarization of the conduction electrons. This inhibition of spin injection (or equally spin flipping of conduction electrons at the interface) results in disappearance of the current.

As indicated by the scale of the color plots, when the Rashba and Dresselhaus spin-orbit interactions compete,
the amplitude of spin density is greatly increased, whereas current density is decreased. The dependence of spin and current densities on ratio $\alpha/\beta$ is discussed in Sec.~\ref{sec:MIamp}.

\subsection{Case of $\alpha/\beta=3$}
\label{sec:aob3}

\begin{figure*}[tb]
\centering
\includegraphics[width=175mm]{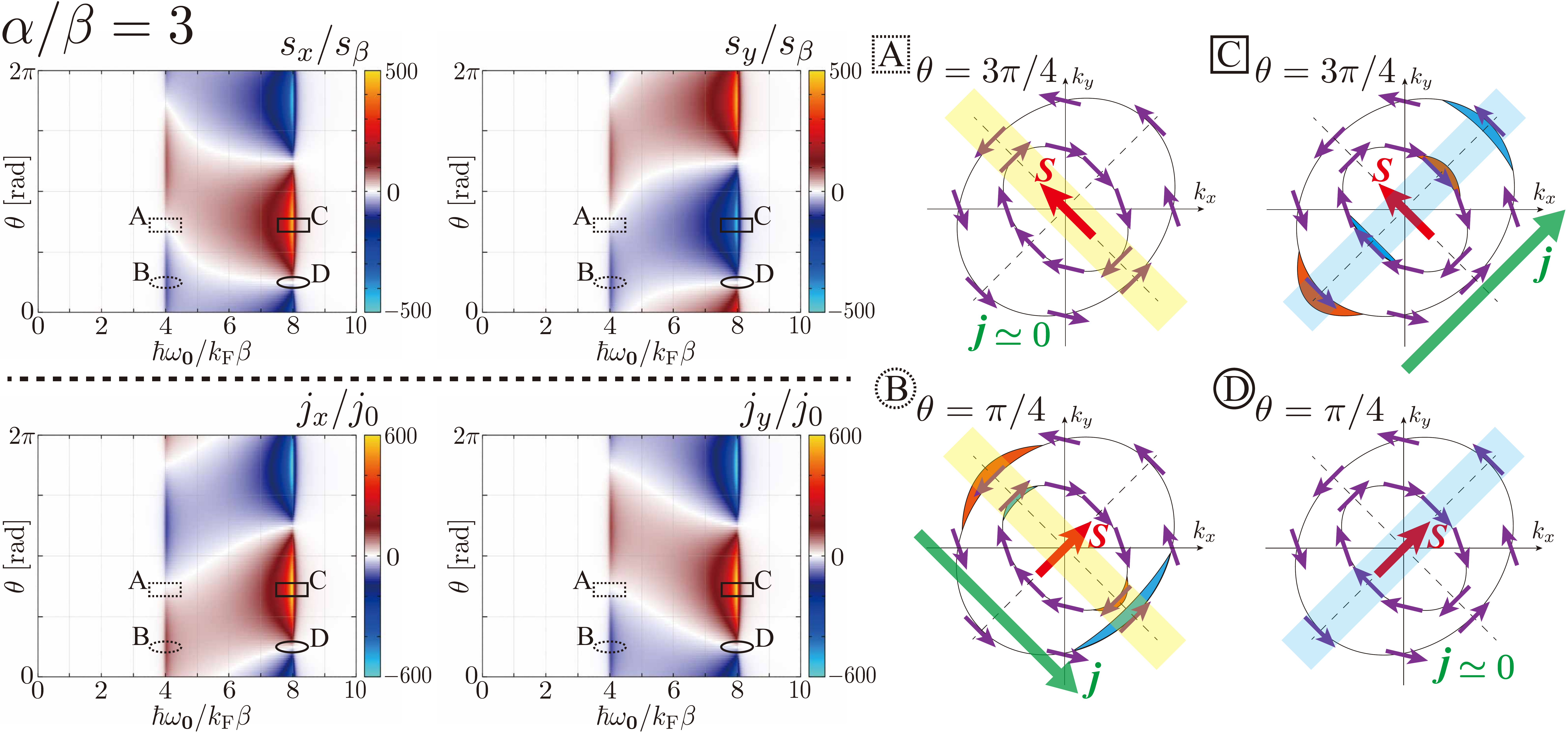}
\caption{Color plots of the spin density $\bm{s}=(s_x, s_y)$ (the upper two panels) and the current density $\bm{j}=(j_x, j_y)$ (the lower two panels) for the case of $\alpha/\beta=3$. The four right insets schematically show the change of the distribution function and the direction of the current in the four regions indicated in the four left panels labelled A, B, C, and D. It was assumed that $\Gamma/k_{\rm F}\beta=0.1$.}
\label{fig:aob3}
\end{figure*}

The final case, namely, the Rashba and Dresselhaus spin-orbit interactions are comparable but have different amplitudes, is discussed next. As an illustrative example, the case of $\alpha/\beta=3$ is considered. Effective Zeeman field $h_{\rm eff}(\varphi)$ depends on $\varphi$ and changes in the range of $2 k_{\rm F}\beta \lesssim h_{\rm eff}(\varphi) \lesssim 4 k_{\rm F}\beta$. As shown in the color plots in Fig.~\ref{fig:aob3}, spin density $\bm{s}/s_\beta=(s_x/s_\beta,s_y/s_\beta)$ and current density $\bm{j}/j_{0}=(j_x/j_{0},j_y/j_{0})$ are induced in the range of $4k_{\rm F}\beta\lesssim \hbar\omega_{\bm{0}}\lesssim 8 k_{\rm F}\beta$, reflecting the distribution of spin splitting $2h_{\rm eff}(\varphi)$. Note that the  dependence of the spin and current densities on $\theta$ differs in the cases of $\hbar\omega_{\bm{0}}\simeq 4k_{\rm F}\beta$ and $\hbar\omega_{\bm{0}}\simeq 8k_{\rm F}\beta$.

Regions A, B, C, and D in the color plots in Fig.~\ref{fig:aob3} are discussed hereafter. Corresponding to these four regions, the four schematic insets on the right side of Fig.~\ref{fig:aob3} are shown for intuitive understanding. In region A, spin injection through the interface can occur only in the yellow region, where spin-splitting energy is smallest. In the yellow region, the spin in the FI is orthogonal to the spin polarization in the 2DEG, so spin injection cannot occur. This situation leads to disappearance of $\bm s$ and ${\bm j}$. On the contrary, in region B, spin-injection rate becomes a maximum since the spin in the FI is parallel to the spin polarization in the 2DEG. Therefore, ${\bm s}$ and ${\bm j}$ take maxima in region B, and the current flows in the direction of $\varphi = 7\pi/4$. 
A similar discussion applies to regions C and D. Spin injection from the interface can occur only in the blue region, in which the spin splitting is largest. In region C (D), spin-injection rate becomes a maximum (zero) since the spin in the FI is parallel (perpendicular) to the spin polarization in the 2DEG, leading to the maximum (minimum) of ${\bm s}$ and ${\bm j}$.

It was also found that the direction of the current rotates as microwave frequency changes. Current ${\bm j}$ flows in the direction of $\varphi=7\pi/4$ for $\hbar \omega_{\bm 0}/k_{\rm F}\beta = 4$, while it flows in the direction of $\varphi = \pi/4$ for $\hbar \omega_{\bm 0}/k_{\rm F}\beta = 8$. This finding can be explained by spin splitting and spin texture on the Fermi surface.

\subsection{Dependence on $\alpha/\beta$}
\label{sec:MIamp}

\begin{figure}[tb]
\centering
\includegraphics[width=70mm]{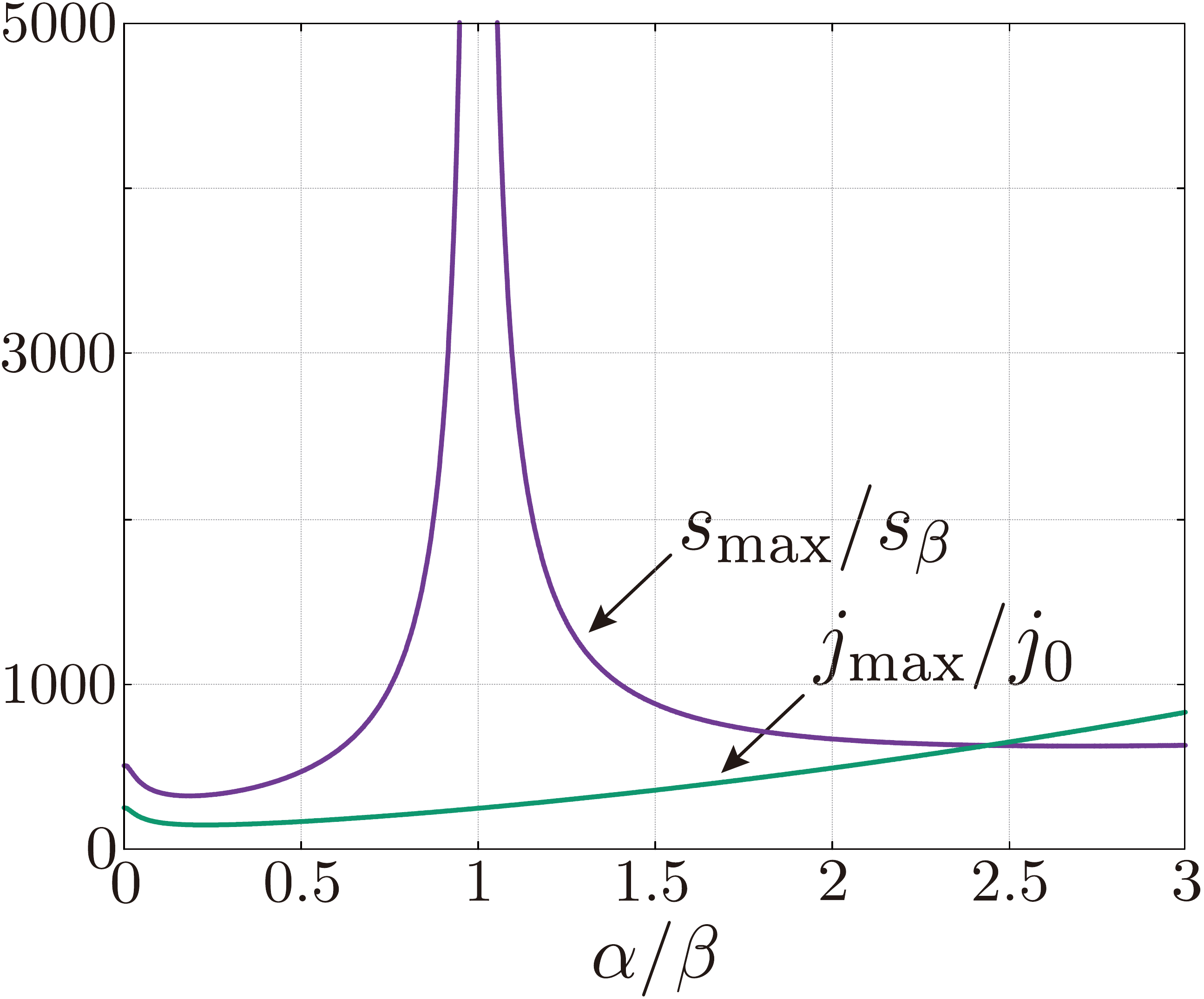}
\caption{Maximum spin density $s_{\rm max}\equiv\max(\sqrt{s_x^2+s_y^2})$ and maximum current density $j_{\rm max}\equiv\max(\sqrt{j_x^2+j_y^2})$ are plotted as a function of $\alpha/\beta$. It was assumed that $\Gamma/k_{\rm F}\beta$=0.1.}
\label{fig:ImaxMmax}
\end{figure}

Dependence of spin and current densities on $\alpha/\beta$ is discussed hereafter. Maximum amplitudes of $\bm{s}$ and $\bm{j}$ are shown as functions of $\alpha/\beta$ in Fig.~\ref{fig:ImaxMmax}. As $\alpha/\beta$ approaches unity, $s_{\rm max}$ diverges, while $j_{\rm max}$ does not show singular behavior there. The former trend reflects the fact that spin-relaxation time becomes substantially long near $\alpha/\beta=1$ because effective Zeeman field ${\bm h}_{\rm eff}$ points in almost the same direction\cite{Bernevig2006, Weber2007, Koralek2009, Kohda2012, Sasaki2014,Yama2023} (see Fig.~\ref{fig:spintexture}~(b)). On the contrary, $j_{\rm max}$ does not diverge at $\alpha/\beta=1$, reflecting cancellation between the contributions from the majority-spin and minority-spin bands.

\section{Experimental Relevance}
\label{Sec:Experiment}

To observe the inverse Rashba-Edelstein effect induced by spin pumping discussed in our work, the weak impurity condition, $\Gamma\ll 2k_{\rm F}\alpha, 2k_{\rm F} \beta \ll \epsilon_{\rm F}$, have to be satisfied, where $2k_{\rm F}\alpha$ or $2k_{\rm F} \beta$ represents the spin-splitting energy near the Fermi surfaces. 
As an example, let us consider a two-dimensional electron gas in AlGaAs/GaAs heterostructures with a high electron mobility of order of $10^{6}\,{\rm cm^2/Vs}$~\cite{Umansky1997}.
In this system, the scattering rate is estimated as $\Gamma \simeq 10^{-1} (2k_{\rm F}\alpha) \simeq 10^{-3}
\epsilon_{\rm F}$~\cite{Yama2021,Yama2023} assuming the electron density $n=5\times 10^{11}~{\rm cm^{-2}}$ and the Rashba spin-orbit interaction $k_{\rm F} \alpha = 0.1~{\rm meV}$ ($\simeq 25 \, {\rm GHz}$)~\cite{Umansky1997,Miller2003}.
Since this estimate satisfies the weak impurity condition well,
we expect that the inverse Rashba-Edelstein effect due to spin pumping can be observed if a junction with FI is fabricated.

As another example, we can consider a thin film of GaAs, where a small number of channels in the thickness direction contribute transport properties.
While YIG/GaAs junctions have recently attracted attention in the spintronics field~\cite{Sadovnikov2019, Nikitov2020, Barman2021}, the inverse Rashba-Edelstein effect has not been studied experimentally yet.
On the other hand, there are experimental studies for Fe/GaAs junctions~\cite{Chen2016}, where the magnitude of the Rashba spin-orbit interaction is about $100~{\rm meV\AA}$~\footnote{We expect that our prediction for the FI-2DEG junctions can apply also for ferromagnetic metal-2DEG junctions qualitatively.
Detailed discussion will be given elsewhere.}. 
Combining this value with the electron density of $n=10^{17}~{\rm cm^{-3}}$ and the electron mobility $\mu = 10^4~{\rm cm^2/Vs}$ at liquid nitrogen temperature in bulk GaAs\cite{Rode1971,Rode1975,Levinshtein1996,Kindyak2002}, the scattering rate is estimated as $\Gamma \simeq 0.5 (2k_{\rm F}\alpha) \simeq 0.1 \epsilon_{\rm F}$.
This estimate indicates that high-quality YIG/GaAs junction may meet the weak impurity condition. 

\section{Summary}
\label{sec:summary}

The inverse Rashba-Edelstein effect (IREE) induced by spin pumping from a ferromagnetic insulator (FI) into a two-dimensional electron gas (2DEG) in which Rashba and Dresselhaus spin-orbit interactions coexist was theoretically investigated. Using the Boltzmann equation in the case that impurity scattering is much weaker than spin-splitting energy in the 2DEG, spin and current densities in the 2DEG caused by the IREE were calculated. It was clarified that the spin and current densities depend on the frequency, $\omega_{\bm 0}$, of the ferromagnetic resonance (FMR) and the azimuth angle, $\theta$, of the spontaneous spin polarization of the FI. It was found that these results are well explained by change of the distribution function of electrons in the spin-splitting bands. It was also found that only the magnitude of spin density increases substantially as the ratio of the magnitudes of the Rashba and Dresselhaus spin-orbit interactions approaches unity, while current density remains finite there. These results can be applied for determining spin textures on the Fermi surface in a 2DEG and would be helpful for understanding and designing of spintronic devices utilizing IREE in 2DEG systems.

In this study, as in previous studies\cite{Yama2021, Yama2023}, a simple parabolic dispersion was considered, and the effect of exchange bias\cite{NOGUES1999203} or band modification due to the interface\cite{Gambardella2011, Rousseau2021} was neglected for simplicity. It is straightforward to extend our formulation of IREE induced by spin pumping to other physical systems with complex band structures and modification by interfacial exchange coupling. We leave such an extended analysis for materials as a future problem.

\section*{Acknowledgements}

The authors thank Y. Suzuki, Y. Kato, M. Kohda, K. Hosokawa, and A. Shitade for their helpful discussions. In particular, we thank Y. Suzuki for his invaluable comments regarding the precise formulation of the induced current. M. Y. was supported by JST SPRING (Grant No.~JPMJSP2108). M.M. was supported by the Priority Program of Chinese Academy of Sciences under Grant No.~XDB28000000, and by JSPS KAKENHI for Grants (No.~JP21H04565, No.~JP21H01800, and No.~JP23H01839) from MEXT, Japan. T. K. acknowledges support from the Japan Society for the Promotion of Science (JSPS KAKENHI Grant No.~JP20K03831). 

\appendix

\section{Derivation of collision terms}
\label{app:CollisionTerms}

The final forms of the collision terms given in Eqs.~(\ref{eq:colpumfin}) and (\ref{eq:goldenimp}) are derived as follows.

The collision term due to spin pumping, given by Eq.~(\ref{eq:colpumfin}), is derived first. Substituting Eqs.~(\ref{eq:fexpan}) and (\ref{eq:Pkkp}) into Eq (\ref{eq:colpump1}) gives 
\begin{align}
&\frac{\partial f(\bm{k},\gamma)}{\partial t}\biggl{|}_{\rm pump}\simeq 
\frac{\pi S_0|\mathcal{T}_{\bm{0}}|^2\langle N_{\bm{0}}\rangle}{\hbar}\nonumber\\
&\times\sum_{\gamma'=\pm}
\sum_{\underset{=\uparrow,\downarrow}{\sigma_1,\sigma_2,\sigma_3,\sigma_4}}
C^*_{\sigma_1 \gamma'}(\varphi)C_{\sigma_2 \gamma}(\varphi)C_{\sigma_3 \gamma'}(\varphi)C^*_{\sigma_4 \gamma}(\varphi)\nonumber \\
&\times\Bigl{[}
(\hat{\sigma}^{x'+})_{\sigma_1\sigma_2}(\hat{\sigma}^{x'+})^*_{\sigma_3\sigma_4}
[f_0(E^{\gamma}_{\bm{k}}-\hbar\omega_{\bm{0}})-f_0(E^{\gamma}_{\bm{k}})]\nonumber \\
&\hspace{15mm}\times \delta((\gamma'-\gamma)h_{\rm eff}(\varphi)+\hbar\omega_{\bm{0}})\nonumber \\
&\hspace{5mm}
+(\hat{\sigma}^{x'-})_{\sigma_1\sigma_2}(\hat{\sigma}^{x'-})^*_{\sigma_3\sigma_4}
[f_0(E^{\gamma}_{\bm{k}}+\hbar\omega_{\bm{0}})-f_0(E^{\gamma}_{\bm{k}})]\nonumber \\
&\hspace{20mm}\times \delta((\gamma'-\gamma)h_{\rm eff}(\varphi)-\hbar\omega_{\bm{0}})
\Bigl{]}. \label{eq:colpC}
\end{align}
Note that $\delta\mu(\varphi,\gamma)$ does not appear in Eq.~(\ref{eq:colpC}) because only up to the first order of $\langle N_{\bm{0}}\rangle$ is considered, and the product of $\langle N_{\bm{0}}\rangle$ and $\delta\mu(\varphi,\gamma)$ is the second order of it. Executing the summation over the spin variables and using Eqs.~(\ref{eq:defC}), (\ref{eq:Cupdown}), and (\ref{eq:hatsig}) make it possible to obtain Eq.~(\ref{eq:colpumfin}).

Applying a random average over the impurity sites to Eq.~(\ref{eq:colimp1}) gives
\begin{align}
&\frac{\partial f(\bm{k},\gamma)}{\partial t}\biggl{|}_{\rm imp}\nonumber \\
&=\frac{2\pi u^2 n_{\rm imp}}{\hbar\mathcal{A}}\sum_{\bm{k}',\gamma'}\sum_{\sigma,\sigma'}
C^*_{\sigma\gamma'}(\varphi')C_{\sigma\gamma}(\varphi)C_{\sigma'\gamma'}(\varphi')C^*_{\sigma'\gamma}(\varphi)
\nonumber \\
&\hspace{5mm}\times [f(\bm{k}',\gamma')-f(\bm{k},\gamma)]\delta(E^{\gamma'}_{\bm{k}'}-E^{\gamma}_{\bm{k}}), \label{eq:impCC}
\end{align}
where $n_{\rm imp}$ is impurity density. Calculating Eq.~(\ref{eq:impCC}) with Eqs.~(\ref{eq:defC}) and (\ref{eq:Cupdown}) gives
\begin{align}
&\frac{\partial f(\bm{k},\gamma)}{\partial t}\biggl{|}_{\rm imp}
=\frac{\pi u^2 n_{\rm imp}}{\hbar\mathcal{A}}
\sum_{\bm{k}',\gamma'}[1+\gamma\gamma'\hat{\bm{h}}_{\rm eff}(\varphi)\cdot\hat{\bm{h}}_{\rm eff}(\varphi')]\nonumber \\
&\hspace{5mm}\times
[f(\bm{k}',\gamma')-f(\bm{k},\gamma)]\delta(E^{\gamma'}_{\bm{k}'}-E^{\gamma}_{\bm{k}}), \label{eq:appgoldenimp}
\end{align}
Additionally, substituting Eq.~(\ref{eq:fexpan}) into Eq.~(\ref{eq:appgoldenimp}) and executing the summation over ${\bm k}'$ with Eq.~(\ref{eq:sumint}) gives Eq.~(\ref{eq:goldenimp}).

\section{Off-diagonal terms of distribution function matrix}
\label{app:off-diagonal}

The Keldysh-Green's function method can be used to derive the following Boltzmann equation\cite{Shytov2006, Suzuki2023} in Hamiltonian $H_{\rm kin}+H_{\rm imp}$ up to the first order of $\bm{h}_{\rm eff}(\varphi)$:
\begin{align}
&
\frac{\partial \hat{f}_{\bm{k}}}{\partial t}
+\frac{1}{2}\Bigl{\{}
\frac{\partial (\xi_{\bm{k}}
-\bm{h}_{\rm eff}(\varphi)\cdot\hat{\bm{\sigma}})}{\partial \bm{k}},\frac{\partial \hat{f}_{\bm{k}}}{\partial \bm{x}}\Bigl{\}}
-
\frac{i}{\hbar}[\bm{h}_{\rm eff}(\varphi)\cdot\hat{\bm{\sigma}},\hat{f}_{\bm{k}}]
\nonumber \\
&=\frac{n_{\rm imp}u^2 \pi}{4}
\sum_{\gamma,\gamma'=\pm}
\int\frac{d^{2}\bm{k}'}{(2\pi)^{2}}
\biggl{(}
2(\hat{f}_{\bm{k}'}-\hat{f}_{\bm{k}})\nonumber \\
&\hspace{5mm}-\{\gamma\hat{\bm{h}}_{\rm eff}(\varphi)\cdot\hat{\bm{\sigma}}
+\gamma'\hat{\bm{h}}_{\rm eff}(\varphi')\cdot\hat{\bm{\sigma}}
,\hat{f}_{\bm{k}'}-\hat{f}_{\bm{k}}\}
\biggl{)}
\delta(E^{\gamma}_{\bm{k}}-E^{\gamma'}_{\bm{k}'})\nonumber \\
&\simeq 
2\pi n_{\rm imp}u^2
\int\frac{d^{2}\bm{k}'}{(2\pi)^{2}}
\biggl{(}
(\hat{f}_{\bm{k}'}-\hat{f}_{\bm{k}})
\cdot\delta(\xi_{\bm{k}}-\xi_{\bm{k}'})\nonumber \\
&\hspace{5mm}-\frac{1}{2}\Bigl{\{} [\bm{h}_{\rm eff}(\varphi)
-\bm{h}_{\rm eff}(\varphi')]\cdot\hat{\bm{\sigma}},\hat{f}_{\bm{k}'}-\hat{f}_{\bm{k}} \Bigl{\}}
\delta'(\xi_{\bm{k}}-\xi_{\bm{k}'})
\biggl{)},\label{eq:Boltzh1}
\end{align}
where $\xi_{\bm{k}} \equiv \epsilon_{\bm{k}}-\mu$, and 
\begin{align}
\hat{f}_{\bm{k}}=
\begin{pmatrix}
f^{\uparrow\uparrow}_{\bm{k}} & f^{\uparrow\downarrow}_{\bm{k}} \\ f^{\downarrow\uparrow}_{\bm{k}} & f^{\downarrow\downarrow}_{\bm{k}}
\end{pmatrix},
\end{align}
is the distribution function of a $2\times 2$ matrix taking the spin into account.
In the last line in Eq.~(\ref{eq:Boltzh1}), the following approximation with $\delta'(x)\equiv d\delta(x)/dx$ is used:
\begin{align}
&\delta(E^{\gamma}_{\bm{k}}-E^{\gamma'}_{\bm{k}'})
= \delta(\xi_{\bm{k}}+\gamma h_{\rm eff}(\varphi)-\xi_{\bm{k}'}-\gamma' h_{\rm eff}(\varphi'))\nonumber \\
&\simeq \delta(\xi_{\bm{k}}-\xi_{\bm{k}'})
+[\gamma h_{\rm eff}(\varphi)-\gamma' h_{\rm eff}(\varphi')]\delta'(\xi_{\bm{k}}-\xi_{\bm{k}'}).
\end{align}
The distribution function for basis $|\bm{k},\gamma=\pm\rangle$ can be written as
\begin{align}
&U^{\dagger}(\varphi)\hat{f}_{\bm{k}}U(\varphi)
=
U^{\dagger}(\varphi)
\begin{pmatrix}
f^{\uparrow\uparrow}_{\bm{k}} & f^{\uparrow\downarrow}_{\bm{k}} \\ f^{\downarrow\uparrow}_{\bm{k}} & f^{\downarrow\downarrow}_{\bm{k}}
\end{pmatrix}
U(\varphi)\nonumber \\
&\equiv\begin{pmatrix}
f^{++}_{\bm{k}} & f^{+-}_{\bm{k}} \\ f^{-+}_{\bm{k}} & f^{--}_{\bm{k}}
\end{pmatrix}
=\begin{pmatrix}
f(\bm{k},+) & f^{+-}_{\bm{k}} \\ f^{-+}_{\bm{k}} & f(\bm{k},-)
\end{pmatrix},  \label{eq:UdfU}\\
&U(\varphi)
=
\begin{pmatrix}
|\bm{k},+\rangle& |\bm{k},-\rangle
\end{pmatrix} 
=\frac{1}{\sqrt{2}}
\begin{pmatrix}
C(\varphi) & C(\varphi) \\ 1 & -1
\end{pmatrix}. \label{defU}
\end{align}
In the steady state and spatial uniform system, the left-hand side of Eq.~(\ref{eq:Boltzh1}) can be written by using Eqs.~(\ref{eq:UdfU}) and (\ref{defU}) as 
\begin{align}
({\rm LHS})&=-
\frac{i}{\hbar}U^{\dagger}(\varphi)[\bm{h}_{\rm eff}(\varphi)\cdot\hat{\bm{\sigma}},\hat{f}_{\bm{k}}]U(\varphi)\nonumber \\
&=\begin{pmatrix}
0 
& 2ih_{\rm eff}(\varphi)f^{+-}_{\bm{k}}/\hbar
\\ -2ih_{\rm eff}(\varphi)f^{-+}_{\bm{k}}/\hbar
& 0
\end{pmatrix}.
\end{align}
As previously noted in the Supplement of Ref.~\onlinecite{Suzuki2023}, order estimation of the right-hand side of Eq. (\ref{eq:Boltzh1}) using the relaxation-time approximation gives the following equation:
\begin{align}
\frac{2ih_{\rm eff}(\varphi)f^{+-}_{\bm{k}}}{\hbar}
\sim \frac{f^{\gamma\gamma}_{\bm{k}}-f_0(\epsilon_{\bm{k}})}{\tau},
~\frac{f^{+-}_{\bm{k}}}{\tau},
~\frac{f^{-+}_{\bm{k}}}{\tau}, \label{eq:relaxapp3term}
\end{align}
where $\gamma=\pm$ and $\tau$ is the electron-relaxation time. Note that $\Gamma$ can be written as $\Gamma=\hbar/\tau$. Under the weak-scattering condition, namely, $\hbar/\tau\ll {\rm max}(2k_{\rm F}\alpha,2k_{\rm F}\beta) \ll \epsilon_{\rm F}$, the following equation holds\cite{Suzuki2023}:
\begin{align}
\frac{f^{+-}_{\bm{k}}}{\tau}&=\frac{\hbar}{2h_{\rm eff}(\varphi)\tau}\cdot\frac{2h_{\rm eff}(\varphi) f^{+-}_{\bm{k}}}{\hbar}\nonumber \\
&\sim \frac{\hbar}{2k_{\rm F}\alpha\tau}\cdot\frac{2h_{\rm eff}(\varphi) f^{+-}_{\bm{k}}}{\hbar} \ll \frac{2h_{\rm eff}(\varphi)f^{+-}_{\bm{k}}}{\hbar}, \label{eq:relaxapp}
\end{align}
where $h_{\rm eff}(\varphi)\sim k_{\rm F}\alpha$ is approximated. Note that this approximation fails around $\varphi=3\pi/4, 7\pi/4$ when $\alpha$ competes with $\beta$. From Eq. (\ref{eq:relaxapp}), the leading term of (\ref{eq:relaxapp3term}) is  $[f^{\gamma\gamma}_{\bm{k}}-f_0(\epsilon_{\bm{k}})]/\tau$, and the following equation can be obtained\cite{Suzuki2023}:
\begin{align}
&\frac{2ih_{\rm eff}(\varphi)f^{+-}_{\bm{k}}}{\hbar}
\sim \frac{\hbar}{2k_{\rm F}\alpha\tau}\cdot\frac{2k_{\rm F}\alpha[f^{\gamma\gamma}_{\bm{k}}-f_0(\epsilon_{\bm{k}})]}{\hbar}\nonumber \\
&\Rightarrow~\frac{f^{+-}_{\bm{k}}}{f^{\gamma\gamma}_{\bm{k}}-f_0(\epsilon_{\bm{k}})}\sim \frac{\hbar}{2k_{\rm F}\alpha\tau}\ll 1.
\end{align}
Similarly, $f^{-+}_{\bm{k}}/[f^{\gamma\gamma}_{\bm{k}}-f_0(\epsilon_{\bm{k}})]\ll 1$ holds under the weak-scattering condition. Therefore, the off-diagonal terms in Eq.~(\ref{eq:UdfU}) can be ignored, and the diagonal components of Eq.~(\ref{eq:Boltzh1}) agree with the expressions of the impurity-collision term calculated by using Fermi's golden rule in Sec.~\ref{sec:impurity} as noted in the supplement of Ref.~\onlinecite{Suzuki2023}. In this work, Y. Suzuki and Y. Kato also clarified that\cite{Suzuki2023}, even in the tunneling Hamiltonian, when the transition rate of the Hamiltonian is much lower than $2k_{\rm F}\alpha$, the off-diagonal terms in the distribution function matrix are small enough to be ignored, and the expression of the collision term calculated by using Fermi's golden rule is correct. Note that the transition rate of the tunneling Hamiltonian, $\mathcal{T}_{\bm{0}}$, is much lower than $2k_{\rm F}\alpha$ for semiconductor heterostructures such as GaAs/AlGaAs\cite{Yama2023}, and the collision term due to spin pumping was calculated by using Fermi's golden rule in Sec.~\ref{sec:spinpumping}.

\bibliography{ref}

\end{document}